\documentclass[aps,prl,reprint,superscriptaddress,footinbib,twocolumn]{revtex4}
\usepackage{dsfont}
\usepackage{amsmath}
\usepackage{graphicx}
\usepackage{multirow}
\usepackage[latin1]{inputenc}
\usepackage{ulem} 
\usepackage{units} 
\usepackage{version} 
\usepackage{color}
\usepackage{colortbl}
\usepackage{xcolor}  
\usepackage[colorlinks=true,citecolor=black,urlcolor=black,linkcolor=black]{hyperref} 

\newcommand{\eff}{_{\text{eff}}}
\newcommand\stxt[1]{_{\text{#1}}} 
\newcommand{\ms}{\ \text{ms}}

\newcommand{\micron}{\ \mu\text{m}}

\begin{document}


\title{Multimode dynamics and emergence of a characteristic length-scale in a one-dimensional quantum system} 



\author{M.~Kuhnert}

\author{R.~Geiger}

\author{T.~Langen}

\author{M.~Gring}

\author{B. Rauer}
\affiliation{Vienna Center for Quantum Science and Technology, Atominstitut, TU Wien, Stadionallee 2, 1020 Vienna, Austria}

\author{T. Kitagawa}

\author{E. Demler}
\affiliation{Harvard-Massachussets Institute of Technology Center for Ultracold Atoms (CUA), Department of Physics, Harvard University, Cambridge, MA 02138, USA.}

\author{D. Adu Smith}

\author{J. Schmiedmayer}
\email[]{schmiedmayer@atomchip.org}
\affiliation{Vienna Center for Quantum Science and Technology, Atominstitut, TU Wien, Stadionallee 2, 1020 Vienna, Austria}


\date{\today}

\begin{abstract}

We study the  non-equilibrium dynamics of a coherently split one-dimensional (1d) Bose gas by measuring the full probability distribution functions of matter-wave interference.
Observing the system on different length scales allows us to probe the dynamics of excitations on different energy scales, revealing two distinct length-scale dependent regimes of relaxation. We measure the crossover length-scale separating these two regimes and identify it with the prethermalized phase-correlation length of the system.
Our approach enables a direct observation of the  multimode dynamics characterizing one-dimensional quantum systems.

\end{abstract}

\maketitle

\label{par:intro}
The non-equilibrium dynamics of many-body quantum systems and their pathway towards equilibrium is of fundamental importance in vastly different fields of physics.
Open questions appear, for example, in high-energy physics for understanding  quark-gluon plasma \cite{Arrizabalaga2005, Arnold2005,Rebhan2005}, in cosmology for describing preheating of the early-universe \cite{Podolsky2006}, or in the comprehension of relaxation processes in condensed matter systems \cite{Micklitz2011,Marino2012}.

Due to their isolation from the environment and their tunability, ultracold atom systems have triggered many studies of non-equilibrium dynamics in closed interacting quantum systems, with a particular interest drawn towards quantum quenches \cite{CazalillaRigol2010NJP,Polkovnikov2011}.
Important questions are related to  systems where the dynamics is constrained by several constants of motion \cite{Kinoshita2006}, and to the possible description of non-equilibrium states by generalized statistical mechanics ensembles  \cite{Rigol2007PRL,Kollar2011}.

Recently, we reported the experimental observation of prethermalization  in a coherently split one-dimensional (1d) ultracold Bose gas \cite{Gring2012},  made possible by a characterization of the dynamical states through measurements of full distribution functions \cite{Hofferberth2008,Kitagawa2011}.
Prethermalization \cite{Berges2004}  was understood  as the rapid relaxation to a steady state   exhibiting thermal-like properties  but differing from the true thermal equilibrium that is eventually expected to occur on longer time-scales \cite{Kollath2007,Moeckel2008PRL,Eckstein2009,Mathey2010,Barnett2011}.

In this letter, we  study the relaxation process \cite{Bistritzer2007} leading to the prethermalized state  by measuring the full distribution functions (FDFs) of phase and contrast of matter-wave interference. We probe the 1d system on different length scales to investigate its multimode dynamics which reveals  two distinct regimes separated by a characteristic crossover length-scale.
We measure this characteristic length-scale and identify it with the effective thermal phase-correlation length of the prethermalized system.

\begin{figure}
	\centering
	\includegraphics[width=0.48\textwidth]{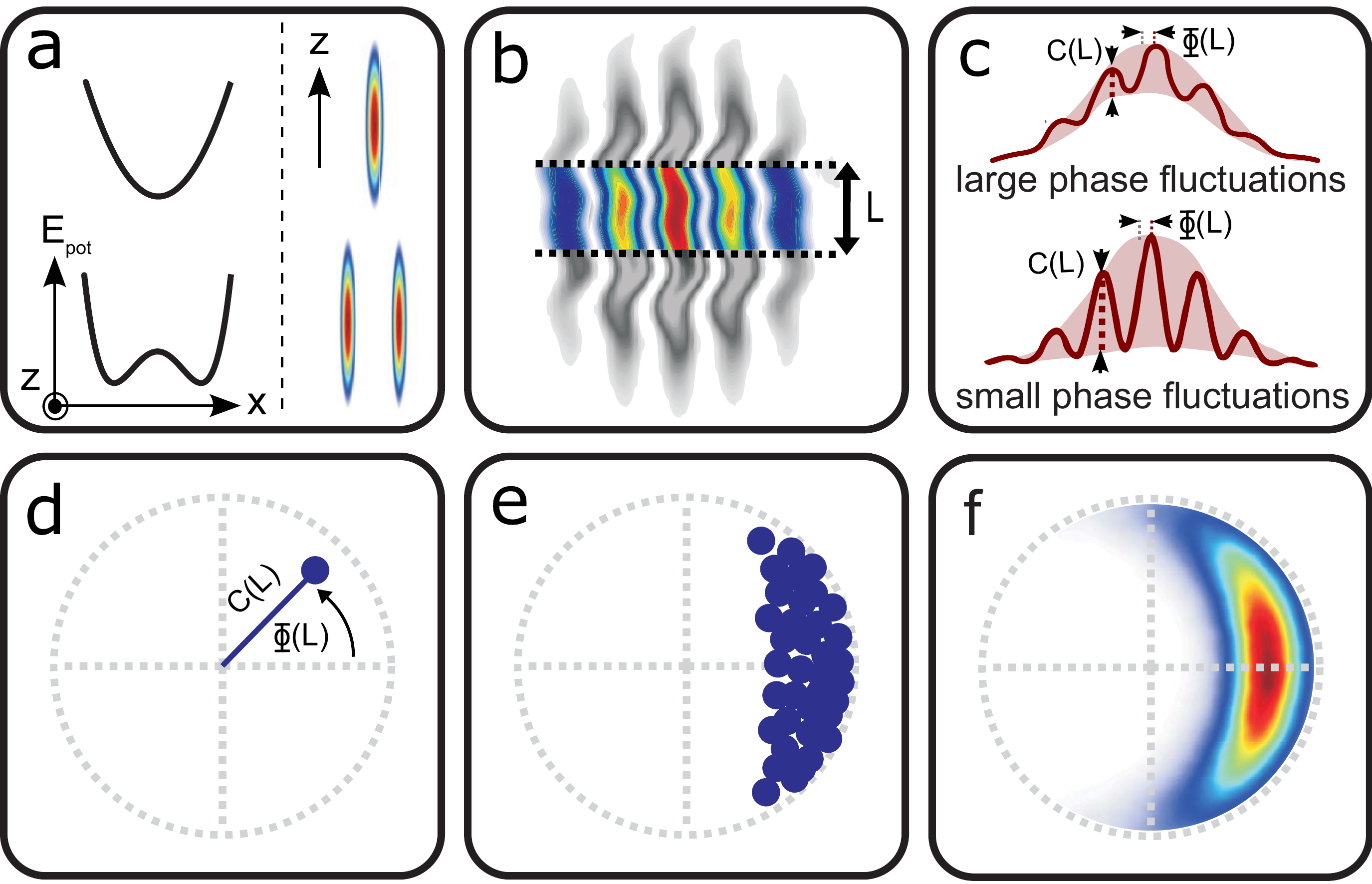} 
	\caption{(Color online) Matter-wave interferometry and correlation measurements. 
	(a) The experiment is initiated by coherently splitting a single quasi-1d Bose gas into two uncoupled gases using a horizontal double-well potential. The system is then held to evolve for a variable evolution time $t_e$, before being released from the trapping potential. 
	The contrast  of the resultant interference pattern is a direct measure of the relative phase fluctuations (b). 
	Integrating the interference pattern over a length $L$ and fitting the resulting line profile (c) gives the phase $\Phi(L)$ and the contrast $C(L)$ which are plotted as circular statistics (d). (e) The whole process is repeated many times (typically 150) to map the FDF for a particular length scale $L$ and evolution time $t_{e}$. The point plot in (e) is then converted into a density plot (f).
	}
	\label{fig:expsetup}
\end{figure}

\begin{figure*}[ht!]
	\centering
		\includegraphics[width=0.7\textwidth]{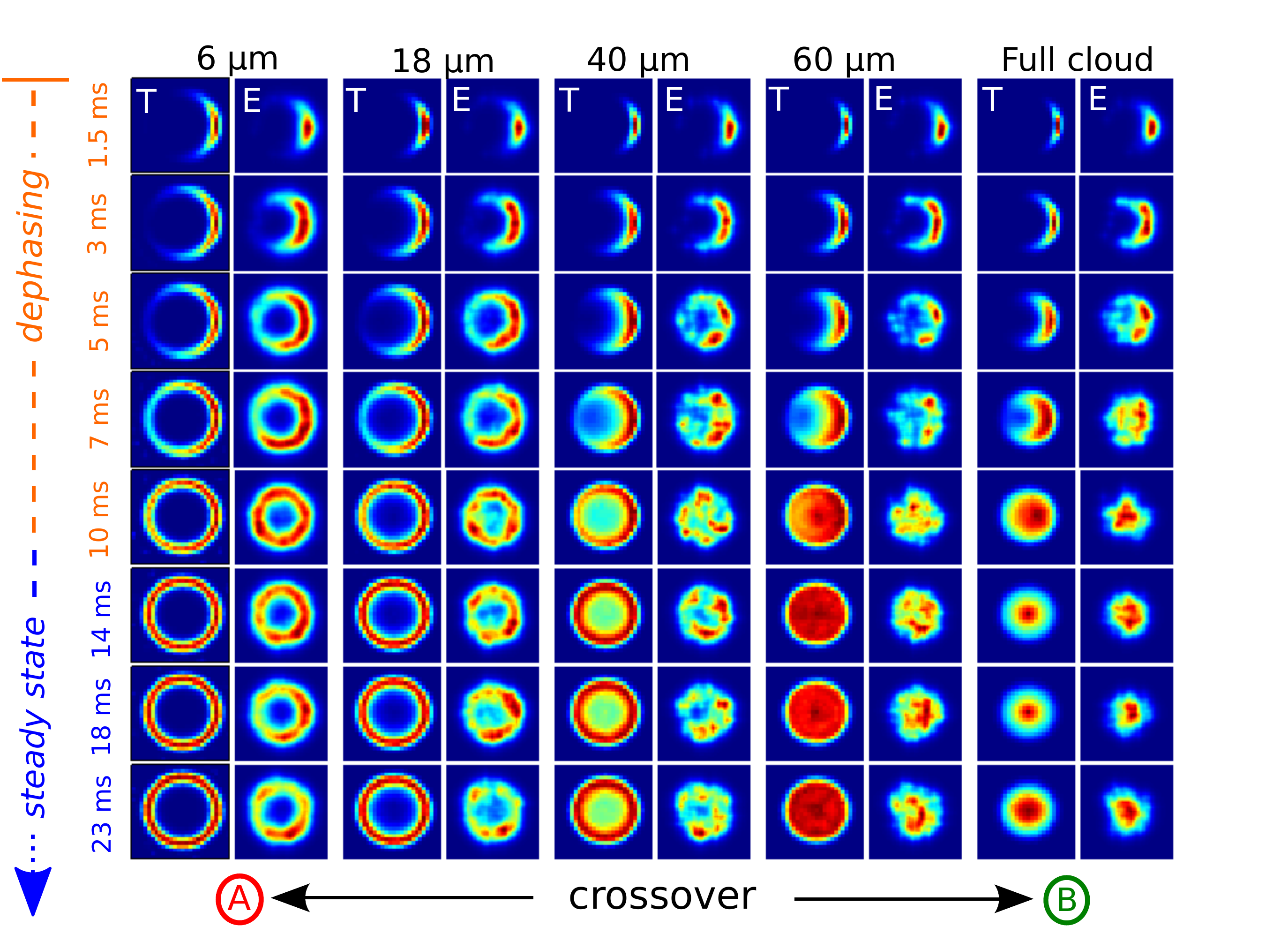}
	\caption{(Color online) Multimode dynamics revealed by FDFs of matter-wave interference.
The probability density of contrast $C(L)$ and phase $\Phi(L)$ of interference patterns is measured for each integration length $L$ (horizontal axis) and evolution time $t_e$ (vertical axis). Red and blue denote high and low probability respectively and the color-map is rescaled for each plot.
For each value of $L$, the right (left) columns correspond to the experimental data (theoretical calculations), and full cloud to $L=100 \micron$.
At $t_e=1.5 \ms$ the high contrasts and small phase spreads demonstrate the coherence of the splitting process.
As time evolves, a steady state emerges and two distinct length-scale dependent regimes appear: the phase diffusion regime (A) and the contrast decay regime (B). 
For short (long) $L$, the phase is random and the probability of observing a high contrast is high (low), resulting in a ring (disk) shape in the density plot.
The theoretical calculations take into account the technical noise on the relative atom number  between the two wells  (standard deviation measured to be $5\%$ of the total atom number) and the error associated with the fitting of the interference patterns.
}
		\label{fig:FDFs}
\end{figure*}

\label{par:exp}  
We prepare a quasi-1d Bose gas of several thousand $^{87}$Rb atoms  in an elongated magnetic microtrap on an atom chip \cite{AtomChips2011} at a (tunable) temperature  between 20 and 120 nK.
The gas is coherently split along the radial direction  using a symmetric radio-frequency dressed-state double-well potential \cite{Schumm2005}, creating two uncoupled 1d gases (Fig.~\ref{fig:expsetup}(a)). The system is then held in the double-well trap for a variable evolution time $t_e$ before being released and allowed to fall under gravity during a time-of-flight of $16 \ms$. 
The resulting matter-wave interference pattern (Fig.~\ref{fig:expsetup}(b)) is recorded using absorption imaging \cite{Smith2011}
and contains information about the local relative phase $\phi(z)=\theta_1(z)-\theta_2(z)$ between the two halves  of the system, with $\theta_{1,2}(z)$ being the fluctuating phase profile of each individual gas.  

In 1d systems, spatial fluctuations arising from excitations at different wavelengths strongly affect the physics \cite{Giamarchi2004}, and probing the system on variable lengths $L$  represents a filter for the effects of these excitations \cite{Kitagawa2010}. 
In our experiment, we integrate the interference pattern longitudinally over a length $L$ and extract a line profile,  from which a contrast $C(L)$ and a phase $\Phi(L)$ are obtained (Fig.\,\ref{fig:expsetup}(c)). 
Repeated realizations  allow us to measure the full probability distribution function (FDF)  of phase and contrast for different evolution times $t_e$ and integration lengths $L$. 
Our measured FDFs are presented in Fig.~\ref{fig:FDFs}.

\label{par:description_results}
For the shortest evolution time (1.5 ms in Fig.~\ref{fig:FDFs}), we observe high contrasts and small phase spreads for all lengths $L$, which demonstrates the coherence of our splitting process. 
After  splitting, the phase fluctuations corresponding to the different excitations grow in magnitude, resulting in a scrambling of the relative phase field $\phi(z)$.
These spatial fluctuations of $\phi(z)$ manifest itself in a randomization of the phase in the interference patterns, and in a decrease of the contrast which depends on the probed length-scale $L$.
For short $L$, the sparsely populated high momentum modes satisfying $k>2\pi/L$ do not lead to contrast  reduction, resulting in a ring-like shape of the FDFs; this is the phase-diffusion regime (A in Fig.~\ref{fig:FDFs}). 
For long  $L$, many modes satisfying $k>2\pi/L$ are populated and their dynamics leads to a scrambling of $\phi(z)$ \textit{within} the probed integration length, resulting in a significant reduction of the probability of observing a high contrast and in a disk-like shape of the FDFs; this is the contrast-decay regime (B).

At longer evolution times ($t_e>10 \ms$), the FDFs do not change significantly any more, revealing the emergence of a steady state characterized by a crossover length-scale between the phase-diffusion (A) and contrast-decay (B) regimes.
In our previous work  \cite{Gring2012}, we identified this steady state as a prethermalized state.

\label{par:theory}
We now quantitatively analyse our observations in the framework of an integrable Luttinger Liquid (LL) theory.
To  describe the dynamics, we consider the LL Hamiltonian $\hat{H} = \frac{\hbar c}{2} \int_{-L/2}^{L/2} dz \left[\frac{K}{\pi}(\nabla \hat{\phi})^{2}+\frac{\pi}{K}\hat{n}^{2}(z)\right]$ 
determining the time evolution of the operators $\hat{\phi}(z)$ and $\hat{n}(z)$ representing the relative phase  and the relative density  of the system, respectively \cite{Bistritzer2007,Kitagawa2010}.
Here, $c=\sqrt{\rho g/m}$ is the sound velocity, $K=\pi\xi\rho$ the LL parameter characterising the strength of the interactions and $\xi=\hbar/m c$ the healing length; $\rho$ is the 1d  density in each half of the system, $g=2\hbar \omega_{\perp} a_s$ the 1d coupling constant, $m$ the mass of the $^{87}$Rb atom, $a_s$ the scattering length and $\omega_{\perp}$ the radial trapping frequency \cite{SupplMat}.
The time evolution of $\hat{n}(z)$ and $\hat{\phi}(z)$ can be described in Fourier space by a set of uncoupled harmonic oscillators of collective modes with momentum $k$ and energies $\hbar\omega_k\approx \hbar c |k|$, where the LL Hamiltonian is diagonal  \cite{Kitagawa2011}.

\label{par:initial_state}
The rapid splitting process prepares a coherent superposition of the atoms in the two wells, resulting in high relative density fluctuations $\langle \hat{n}_k^{2} \rangle _{t=0} = \rho/2$ and small relative phase fluctuations $\langle \hat{\phi}_k^{2} \rangle _{t=0} = 1/2\rho$ \cite{SupplMat,Kitagawa2011}.
In the experiment, we observe enhanced global phase fluctuations for $t_e=1.5 \ms$  with respect to quantum noise, which are attributed to technical noise on the relative atom number between the two wells. 
For the dynamics of the contrast, the contribution of this technical noise  is however negligible compared to that of the initial density fluctuations,  for the probed integration lengths ($L\gg \xi$).
Focusing on the contrast thus allows us to directly observe the effect of quantum noise associated with the splitting process through the evolution of the many-body system.


The results of the LL calculations, taking into account our  technical noise, are presented in the left columns  of Fig.~\ref{fig:FDFs}.
For all length scales and evolution times, we find very good agreement between the experimental  and theoretical FDFs. Our approach  gives a direct and intuitive visualization of the multimode nature of the dynamics in 1d Bose gases \cite{Widera2008}.

\begin{figure}[ht!]
	\centering
		\includegraphics[width=0.4\textwidth]{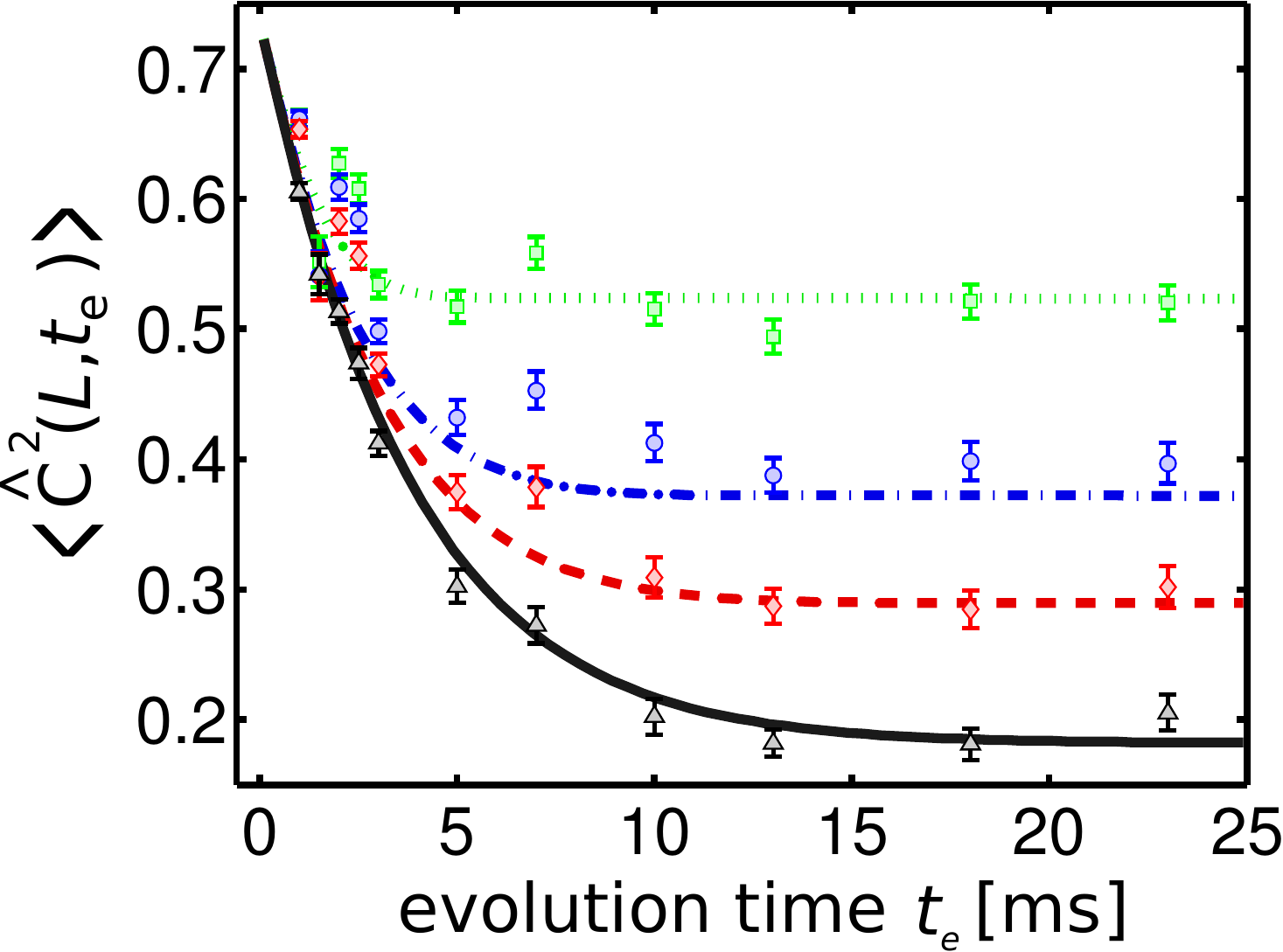}
	\caption{(Color online). Contrast dynamics. Measured values of the mean squared contrast  for various integration lengths, corrected for the contrast reduction factor due to the imperfections of our imaging system \cite{SupplMat}. From top to bottom: $L=$ 18, 40, 60, 100 $\mu m$. The lines show the results of the Luttinger Liquid calculations for these integration lengths.
	The theory data for all $L$ have been rescaled by the \textit{same} factor $r=0.74$ \cite{NoteNormalizationFactor}.  
	Apart from this  factor,	 no fitting parameter is used. 
	We observe a relaxation process in which a steady-state is established on a time-scale depending on $L$  and corresponding to the dephasing of the different excitations probed within that $L$. 
}
	\label{fig:Fig3}
\end{figure}

\label{par:timescale}
To study the relaxation of the contrast in more detail,  we compare in Fig.~\ref{fig:Fig3}  the measured mean values of the integrated squared contrast $ \langle \hat{C}^2(L,t_e)\rangle$ to the prediction of the LL theory, taking into account the imperfections of our imaging system \cite{SupplMat}.
We  observe that the characteristic time for the decay of the contrast is longer for long $L$ than for short $L$, which is well captured by our model.  
For a more quantitative comparison, we fit the experimental and theory data by an exponential decay.
For the experiment, we  find characteristic decay times $\tau_{\text{exp}}(L=6 \micron)=0.53\pm0.33  \ms$ and $\tau_{\text{exp}}(L=100 \micron)=2.89\pm0.36  \ms$ (standard error on fitted parameter).
Fitting the theoretical calculations for our measured LL parameter $K = 58 \pm 4$, we find $\tau_{\text{th}}(L=6 \micron)=0.40 \pm 0.11 \ms$ and $\tau_{\text{th}}(L=100 \micron)=3.15 \pm 0.21  \ms$, in good agreement with the observations.
The dependence of the relaxation time-scale on $L$ is an additional clear signature of the multimode nature of the  dynamics \cite{Kitagawa2011}. 

Fig.~\ref{fig:Fig3} also  reveals the onset of the prethermalized state for $t_e>10 \ms$ which is characterized by high (low) contrasts at short (long) integration lengths, in line with the probability density plots A (B) of Fig.~\ref{fig:FDFs}.
In the rest of this letter, we investigate the crossover between the  high contrasts and  low contrasts  regimes observed in Figs.~\ref{fig:FDFs}-\ref{fig:Fig3}.

\begin{figure}[ht!]
	\centering
		\includegraphics[width=0.48\textwidth]{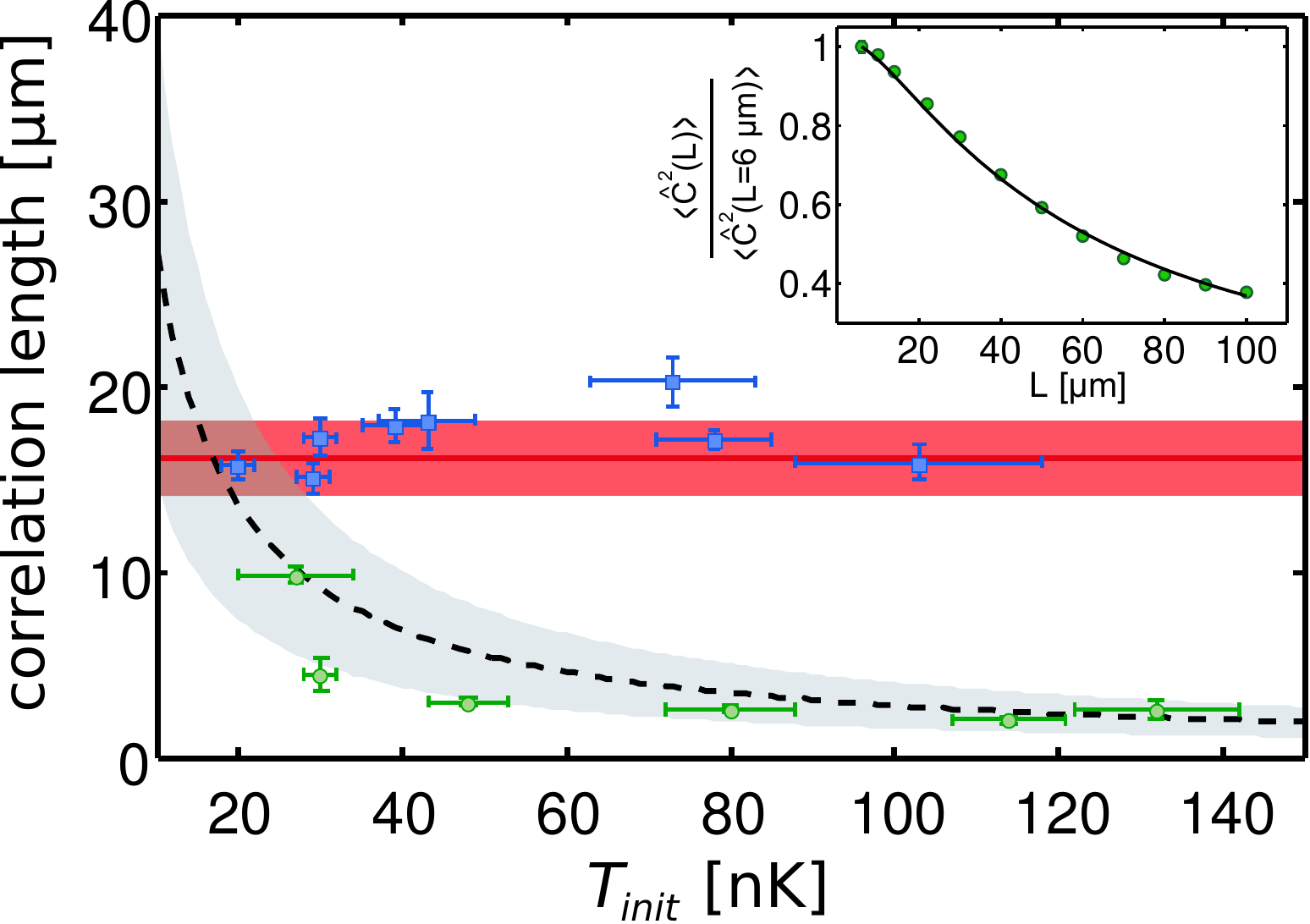}
	\caption{(Color online) Comparison of the relative-phase correlation length of the prethermalized system ($\lambda\eff$) and of a system at thermal equilibrium ($\lambda_{\phi}$).
	 Blue squares: values of $\lambda\eff$  obtained by varying the temperature $T\stxt{init}$ before splitting the quasi-condensate (see main text). 
	  	Error bars denote one standard deviation and are obtained by a bootstrapping method  \cite{SupplMat}.
	The red shaded area corresponds to the $95\%$ confidence interval around the theoretical calculation (line) accounting for the uncertainty on the  experimental parameters. 
	Green circles:  correlation length $\lambda_{\phi}$ of a system of two quasi-condensates in thermal equilibrium, normalized to the mean of the different densities. Gray shaded area: $95\%$ confidence interval around the theory (dashed line).
	Inset: estimation of the correlation length. The filled circles are the values of $\langle\hat{C}^2(L)\rangle$  in the prethermalized state (data of Fig.~\ref{fig:Fig3} for $t_e>10 \ms$),  normalized to the value at $L=6 \ \mu m$.  The solid line is a fit to an effective thermal equilibrium theory with  the correlation length being the only free parameter.
	}
	\label{fig:Fig4}
\end{figure}

\label{par:crossover}
Within the LL theory, the crossover length-scale separating the phase diffusion and contrast decay regimes can be calculated analytically \cite{Kitagawa2011}: $l_{0}  = 2\hbar^2 / m g = \hbar/m a_s \omega_{\perp}$.
For the data in Figs.~\ref{fig:FDFs}-\ref{fig:Fig3} we measure $\omega_{\perp}/2\pi=1.40 \pm 0.08 \ \text{kHz}$ and calculate $ l_0 =  15.8\pm0.9 \, \mu$m.
To understand the nature of this crossover length-scale, we consider the decay of the mean squared contrast $\langle\hat{C}^2(L)\rangle$ as a function of $L$ for $t_e>10 \ms$ (inset of Fig.~\ref{fig:Fig4}). 
To fit this decay, we use our knowledge of the correlation functions in the prethermalized state, which have been shown to be thermal-like  \cite{Kitagawa2011,Gring2012}. 
Using the direct link between the mean squared contrast and the relative-phase correlation function \cite{Polkovnikov2006,SupplMat},  a fit  of $\langle\hat{C}^2(L)\rangle$ allows us to estimate the effective thermal phase-correlation length $\lambda\eff$ of the system.
For the data in Fig.~\ref{fig:Fig4} (inset) we find $\lambda\eff=16.6 \pm 0.9 \ \mu m$, close to the  value of $l_0$ calculated above using the independently determined experimental parameters. 
Note that our method of estimating the correlation length does not require precise measurements of the system parameters to fit the FDFs, but only needs the knowledge that the correlation functions are effectively thermal.

The equivalence of the crossover length-scale $l_0$ and the effective thermal phase-correlation length $\lambda\eff$  can be understood in the following way. 
In the case of two uncoupled quasi-condensates in thermal equilibrium with temperature $T$, the thermal phase-correlation length is given by $\lambda_{\phi}=\hbar^2\rho/m k\stxt{B} T$ \cite{Bouchoule2003,Betz2011}.
Due to our rapid splitting process, the  energy initially stored in the system  is equally distributed between the different $k$-modes of the system \cite{Kitagawa2011,Gring2012}, resulting in thermal-like correlations  characterized by an effective temperature $k\stxt{B}T\eff = \langle \hat{H}\rangle|_{t=0}=\rho g/2$.
The effective thermal phase-correlation length can thus be identified with $\lambda\eff=\hbar^2\rho/m k\stxt{B} T\eff=2\hbar^2/m g$, and is equivalent to the crossover length scale $l_0$.

\label{par:scaling_with_T}
While the prethermalized system reveals thermal-like correlations, its correlation length $\lambda\eff$  depends only on the 1d coupling constant $g$, in contrast to a system of two quasi-condensates at thermal equilibrium where  $\lambda_{\phi}$ is a function of density and temperature.
To reveal this difference experimentally, we varied the initial temperature $T\stxt{init}$  of the quasi-condensate before splitting and measured the correlation length by fitting the decay of $\langle\hat{C}^2(L)\rangle$  (see inset of Fig.~\ref{fig:Fig4}). 
The temperature before splitting, $T\stxt{init}$, was obtained through measurements of the second order correlation function of longitudinal density fluctuations after time-of-flight \cite{Manz2010,SupplMat}. The results (blue squares in Fig.~\ref{fig:Fig4}) show the independence of $\lambda\eff$ from $T\stxt{init}$. 
When varying  $T\stxt{init}$, we  further observed that the evolution of the FDFs remained close to that presented in Fig.~\ref{fig:FDFs}. In particular, the steady states reached in the evolution exhibited the same crossover  between the phase-diffusion and contrast-decay regimes.
These observations  confirm our interpretation of the crossover length-scale $l_0$ as the prethermalized phase-correlation length $\lambda\eff$ of the system.
Performing experiments with  two quasi-condensates in thermal equilibrium with temperature $T$ and prepared in the same double-well trap \cite{SupplMat}, we observe  $\lambda_{\phi}\propto 1/ T$ (green circles in Fig.~\ref{fig:Fig4}), emphasizing the  different characteristic scalings of the thermal and prethermalized phase-correlation lengths.

\label{par:discussion}
In conclusion, we demonstrated the multimode nature of the dynamics of a coherently split 1d Bose gas using  matter-wave interferometry. 
Our approach allows the direct observation of the effect of the quantum noise present in the initial non-equilibrium state on the evolution of the many-body  system.
We showed the emergence of two distinct  regimes of relaxation separated by a characteristic crossover length scale, which we identified as the prethermalized phase-correlation length of the system.
This characteristic  length-scale, $2\hbar^2/mg$,  reflects the parametrization of the many-body theory describing the dynamics of the system   by a single  parameter, $g$.
For our closed system to thermalize  and the phase-correlation length to reach the true thermal equilibrium value $\lambda_{\phi}$,  the integrability of the LL theory needs to be broken. 
Investigations into the processes leading to the full thermalization of the system are ongoing.

\paragraph{Acknowledgements.}
\label{par:ack}
We acknowledge fruitful discussions with Igor Mazets.
Our work was supported by the FWF through the Wittgenstein Prize and the EU through the integrating project AQUTE.  
MK, TL and MG thank the FWF Doctoral Programme CoQuS (\textit{W1210}), RG is supported by the FWF through the Lise Meitner Programme M 1423, and DAS acknowledges the EU (grant $N^{o.}\,$\textit{220586}).
TK and ED thank DARPA, Harvard-MIT CUA, NSF (Grant No. DMR-07-05472) and ARO-MURI on Atomtronics.


\bibliography{Kuhnert_main_text_with_figures}




\end{document}